\documentclass[12pt,preprint,longnamesfirst]{aastex}
\usepackage{amsmath}

\newcommand{\bpi}{\boldsymbol\pi}
\newcommand{\btheta}{\boldsymbol\theta}
\newcommand{\bmu}{\boldsymbol\mu}
\newcommand{\e}{\mathrm{E}}
\renewcommand{\max}{{\rm max}}
\newcommand{\mas}{{\rm mas}}
\newcommand{\rel}{{\rm rel}}
\newcommand{\kpc}{{\rm kpc}}
\newcommand{\masyr}{{\rm mas}\,{\rm yr}^{-1}}

\begin{document}

\title{Potential Direct Single-Star Mass Measurement}

\author{
H.~Ghosh\altaffilmark{1},
D.L.~DePoy\altaffilmark{1},
A.~Gal-Yam\altaffilmark{2,3},
B.S.~Gaudi\altaffilmark{4},
A.~Gould\altaffilmark{1},
C.~Han\altaffilmark{1,5},
Y.~Lipkin\altaffilmark{6},
D.~Maoz\altaffilmark{6},
E.~O.~Ofek\altaffilmark{6},
B.-G.~Park\altaffilmark{7},
R.W.~Pogge\altaffilmark{1},
and S.~Salim\altaffilmark{8} \\
(The $\mu$FUN Collaboration) \\ \vspace{\baselineskip}
F.~Abe\altaffilmark{9},
D.P.~Bennett\altaffilmark{10},
I.A.~Bond\altaffilmark{11},
S.~Eguchi\altaffilmark{9},
Y.~Furuta\altaffilmark{9},
J.B.~Hearnshaw\altaffilmark{12},
K.~Kamiya\altaffilmark{9},
P.M.~Kilmartin\altaffilmark{12},
Y.~Kurata\altaffilmark{9},
K.~Masuda\altaffilmark{9},
Y.~Matsubara\altaffilmark{9},
Y.~Muraki\altaffilmark{9},
S.~Noda\altaffilmark{13},
K.~Okajima\altaffilmark{9},
N.J.~Rattenbury\altaffilmark{14},
T.~Sako\altaffilmark{9},
T.~Sekiguchi\altaffilmark{9},
D.J.~Sullivan\altaffilmark{15},
T.~Sumi\altaffilmark{16},
P.J.~Tristram\altaffilmark{14},
T.~Yanagisawa\altaffilmark{16}, and
P.C.M.~Yock\altaffilmark{14}\\
(The MOA Collaboration)\\ \vspace{\baselineskip}
A.~Udalski\altaffilmark{18},
I.~Soszy{\'n}ski\altaffilmark{18},
{\L}.~Wyrzykowski\altaffilmark{18,6},
M.~Kubiak\altaffilmark{18},
M.~K.~Szyma{\'n}ski\altaffilmark{18},
G.~Pietrzy{\'n}ski\altaffilmark{18,19},
O.~Szewczyk\altaffilmark{18},
and K.~\.Zebru{\'n}\altaffilmark{18} \\
(The OGLE Collaboration)
\and
M.~D.~Albrow\altaffilmark{20},
J.-P.~Beaulieu\altaffilmark{21},
J.~A.~R.~Caldwell\altaffilmark{22},
A.~Cassan\altaffilmark{21},
C.~Coutures\altaffilmark{21,23},
M.~Dominik\altaffilmark{24},
J.~Donatowicz\altaffilmark{25},
P.~Fouqu\'e\altaffilmark{26},
J.~Greenhill\altaffilmark{27},
K.~Hill\altaffilmark{27},
K.~Horne\altaffilmark{24},
U.~G.~J\o rgensen\altaffilmark{28},
S.~Kane\altaffilmark{24},
D.~Kubas\altaffilmark{29},
R.~Martin\altaffilmark{30},
J.~Menzies\altaffilmark{31},
K.~R.~Pollard\altaffilmark{20},
K.~C.~Sahu\altaffilmark{22},
J.~Wambsganss\altaffilmark{28},
R.~Watson\altaffilmark{27},
A.~Williams\altaffilmark{30}\\
(The PLANET Collaboration\altaffilmark{32})
}

\altaffiltext{1}
{Department of Astronomy, The Ohio State University,
140 West 18th Avenue, Columbus, OH 43210, USA; depoy, ghosh, gould, 
pogge@astronomy.ohio-state.edu}
\altaffiltext{2}
{Department of Astronomy, MS 105-24, California Institute of Technology,
Pasadena, CA 91025, USA; avishay@astro.caltech.edu}
\altaffiltext{3}
{Hubble Fellow}
\altaffiltext{4}
{Harvard-Smithsonian Center for Astrophysics, Cambridge, MA 02138, USA; 
sgaudi@cfa.harvard.edu}
\altaffiltext{5}
{Department of Physics, Institute for Basic Science Research,
Chungbuk National University, Chongju 361-763, Korea;
cheongho@astroph\-.chungbuk.ac.kr}
\altaffiltext{6}
{School of Physics and Astronomy and Wise Observatory, Tel Aviv University, Tel Aviv 69978, Israel; yiftah@wise.tau.ac.il, dani@wise.tau.ac.il, eran@wise.tau.ac.il}
\altaffiltext{7}
{Korea Astronomy Observatory,
61-1, Whaam-Dong, Youseong-Gu, Daejeon 305-348, Korea; bgpark@boao.re.kr}
\altaffiltext{8}
{Department of Physics \& Astronomy, University of California at Los
Angeles, Los Angeles, CA 90095}

\altaffiltext{9}
{Solar-Terrestrial Environment Laboratory, Nagoya University,
Nagoya 464-8601, Japan; abe, furuta, kkamiya, kmasuda, kurata, 
muraki, okajima, sado, sako, sekiguchi, ymatsu@stelab.nagoya-u.ac.jp}
\altaffiltext{10}
{Department of Physics, Notre Dame University, Notre Dame, IN 46556, USA;
bennett@emu.phys.nd.edu}
\altaffiltext{11}
{Institute for Astronomy, University of Edinburgh, Edinburgh, EH9 3HJ, UK;
iab@roe.ac.uk}
\altaffiltext{12}
{Department of Physics and Astronomy, University of Canterbury,
Private Bag 4800, Christchurch, New Zealand;
john.hearnshaw, pam.kilmartin@canterbury.ac.nz}
\altaffiltext{13}
{National Astronomical Observatory of Japan, Tokyo, Japan;
sachi.t.noda@nao.ac.jp}
\altaffiltext{14}
{Department of Physics, University of Auckland, Auckland, New Zealand;
nrat001@phy.auckland.ac.nz, paulonieka@hotmail.com,
p.yock@auckland.ac.nz}
\altaffiltext{15}
{School of Chemical and Physical Sciences, Victoria University,
PO Box 600, Wellington, New Zealand; 
denis.sullivan@vuw.ac.nz}
\altaffiltext{16}
{Department of Astrophysical Sciences, Princeton University,
Princeton NJ 08544, USA; sumi@astro.princeton.edu }
\altaffiltext{17}
{National Aerospace Laboratory, Tokyo, Japan; tyanagi@nal.go.jp}

\altaffiltext{18}
{Warsaw University Observatory, Al.~Ujazdowskie~4, 00-478~Warszawa,
Poland;
udalski, soszynsk, wyrzykow, mk, msz, pietrzyn, szewczyk,
zebrun@astrouw.edu.pl}
\altaffiltext{19}
{Universidad de Concepci{\'o}n, Departamento de Fisica, Casilla 160-C,
Concepci{\'o}n, Chile}

\altaffiltext{20}
{University of Canterbury, Department of Physics \& Astronomy, Private Bag
4800, Christchurch, New Zealand }
\altaffiltext{21}
{Institut d'Astrophysique de Paris, 98bis Boulevard Arago, 75014 Paris,
France}
\altaffiltext{22}
{Space Telescope Science Institute, 3700 San Martin Drive, Baltimore, MD
21218, USA}
\altaffiltext{23}
{DSM/DAPNIA, CEA Saclay, 91191 Gif-sur-Yvette cedex, France}
\altaffiltext{24}
{University of St Andrews, School of Physics \& Astronomy, North Haugh,
 St Andrews, KY16~9SS, United Kingdom; md35@st-andrews.ac.uk}
\altaffiltext{25}
{Technical University of Vienna, Dept.\ of Computing, Wiedner
Hauptstrasse 10, Vienna, Austria}
\altaffiltext{26}
{Observatoire Midi-Pyrenees, UMR 5572, 14, avenue Edouard Belin, F-31400
Toulouse, France}
\altaffiltext{27}
{University of Tasmania, Physics Department, GPO 252C, Hobart, Tasmania
7001, Australia}
\altaffiltext{28}
{Niels Bohr Institute, Astronomical Observatory, Juliane Maries Vej 30,
DK-2100 Copenhagen, Denmark}
\altaffiltext{29}
{Universit\"at Potsdam, Astrophysik, Am Neuen Palais 10, D-14469 Potsdam,
Germany}
\altaffiltext{30}
{Perth Observatory, Walnut Road, Bickley, Perth 6076, Australia}
\altaffiltext{31}
{South African Astronomical Observatory, P.O. Box 9 Observatory 7935,
South Africa}
\altaffiltext{32}
{email address: planet@iap.fr}

\begin{abstract}

We analyze the lightcurve of the microlensing event
OGLE-2003-BLG-175/MOA-2003-BLG-45 and show that it has two properties
that, when combined with future high resolution astrometry, could lead
to a direct, accurate measurement of the lens mass.  First, the
lightcurve shows clear signs of distortion due to the Earth's
accelerated motion, which yields a measurement of the projected Einstein
radius $\tilde r_\e$.  Second, from precise astrometric measurements, we
show that the blended light in the event is coincident with the
microlensed source to within about 15 mas.  This argues strongly that
this blended light is the lens and hence opens the possibility of
directly measuring the lens-source relative proper motion $\bmu_\rel$
and so the mass $M=(c^2/4G)\mu_\rel t_\e\tilde r_\e$, where $t_\e$ is
the measured Einstein timescale. While the lightcurve-based measurement
of $\tilde r_\e$ is, by itself, severely degenerate, we show that this
degeneracy can be completely resolved by measuring the direction of
proper motion $\bmu_\rel$.

\end{abstract}


\section{Introduction
\label{sec:intro}}

When microlensing experiments were initiated more than a decade ago
\citep{alcock93,aubourg93,udalski93}, there was no expectation that
the individual lens masses could be determined to much better than an
order of magnitude.  The only routinely observable parameter, the
Einstein timescale $t_\e$, is related in a complicated way to the
mass $M$ and two other parameters, the lens-source relative
parallax, $\pi_\rel$, and relative proper motion $\mu_\rel$.
\begin{equation}
t_\e = \frac{\theta_\e}{\mu_\rel},
\label{eqn:tedef}
\end{equation}
where
\begin{equation}
\label{eqn:thetaekappa}
\theta_\e = \sqrt{\kappa M \pi_\rel}
\end{equation}
is the angular radius of the Einstein ring and $\kappa \equiv 4G/(c^2
\mathrm{AU}) \simeq 8.14\,\mathrm{mas}/M_\odot$.  In principle,
therefore, a measurement of $\theta_\e$ and $\pi_\rel$ would lead to a
determination of lens mass \citep{refsdal64}.  However, since neither
$\pi_\rel$ nor $\mu_\rel$ are usually known for microlensing events, one
can generally obtain only a rough estimate of the lens mass based on
statistical inferences from the distance and velocity distributions of
the lens and source populations.

The motion of the Earth in its orbit produces a distortion in the
observed lightcurve from that of the simple heliocentric case. The
magnitude of this distortion is proportional to the size of the
projected Einstein radius $\tilde{r}_\e$ relative to the size of the
Earth's orbit. This ratio, $\pi_\e \equiv \mathrm{AU}/\tilde{r}_\e$, is
commonly called the microlens parallax, from the similarity in its
definition to astrometric parallax. As shown in \citet{gould00}, $\pi_\e
\theta_\e = \pi_\rel$, and therefore
\begin{equation}
\pi_\e = \sqrt{\frac{\pi_\rel}{\kappa M}}.
\label{eqn:pie}
\end{equation}

\citet{gould92} pointed out that individual lens masses could be
determined provided that $\theta_\e$ and $\pi_\e$ were
simultaneously measured for the same event, and he suggested some
methods for measuring each%
\footnote{The relationship between the observable parameters $\theta_\e$
and $\tilde{r}_\e$ and the physical parameters $M$ and $\pi_\rel$ is
explained in \citet{gould00}. See especially his Fig. 1.}.
If successfully carried out, microlensing
would join only a handful of other methods for directly measuring
stellar masses.  However, unlike all other methods, microlensing can in
principle be used to measure the masses of objects without visible
companions, in particular, single stars. At present, the Sun is the only
single star whose mass has been directly measured with high precision.
This was possible originally only because it has non-stellar, but
nevertheless highly visible companions.

In fact, the Sun is also the one single star whose mass has been
accurately measured using gravitational lensing.  While the original
Eddington eclipse experiment was regarded at the time as a confirmation
of Einstein's general relativity \citep{dyson}, general relativity is by
this point so well established that this experiment can now be regarded
as a mass measurement of the Sun.  Applying the same principle to $10^5$
stars in the {\it Hipparcos} catalog, \citet{fma} were able to confirm general
relativity (or alternatively measure the mass of the Sun) accurate to
0.3\%.

With the notable exception of the Sun, and despite the discovery of
several thousand microlensing events as well as a decade of theoretical
efforts to invent new ways to measure $\pi_\e$ and $\theta_\e$, there
have been just two mass measurements of single stars using microlensing.
The problem is that while the microlens parallax $\pi_\e$ has been
measured for more than a dozen single lenses
\citep*{al1,alcock01,Ma99, OGLE, bond01, O9932, OGLEII,
blackhole,smp03,jiang},
the angular Einstein radius $\theta_\e$ has been measured for only
five single lenses \citep{alcock97,alcock01,smw03,yoo,jiang}. 
Though \citet{alcock01} and \citet{jiang} each measured both $\theta_\e$
and $\pi_\e$ for their events, respectively MACHO-LMC-5 and
OGLE-2003-BLG-238, in neither case was the $\pi_\e$ measurement very
accurate. Moreover, \citet{lmc5} showed that the microlens parallax
measurement of MACHO-LMC-5 was subject to a discrete
degeneracy. Nevertheless, \citet{dck04} resolved this degeneracy by a
trigonometric measurement of $\pi_\rel$. \citet{lmc5b} then combined the
\citet{dck04} measurement of $\pi_\rel$ and $\bmu_\rel$ with the
original photometric data and additional high resolution photometry of
the source to constrain the mass to within 17\%. This is the most
precise direct mass measurement of a single star (other than the Sun) to
date.  By comparison, the mass of the only other directly measured
single star, OGLE-2003-BLG-238, is only accurate to a factor of a few.

\citet{jin} made the most precise microlens mass measurement to date,
with an error of just 9\%.  However, the lens, EROS-BLG-2000-5, was a
binary.  In the future, the {\it Space Interferometry Mission} should
routinely measure the masses of single stars both for stars in the bulge
\citep{gssim} and for nearby stars passing more distant ones
\citep{refsdal64,pac95,salimgould}.  Thus, at present, the direct
measurement of single star masses (other than the Sun) remains a
difficult undertaking.

Here we present evidence that the microlensing event
OGLE-2003-BLG-175/MOA-2003-BLG-45 is an excellent candidate for such a
single-star measurement.  This seems odd at first sight because, as we
will show, $\pi_\e$ is measured only to a factor of a few and
$\theta_\e$ is not measured at all.  Hence it would appear difficult to
derive any mass measurement, let alone a precise one.  However, the
event has the relatively unusual property  that the lens itself is
visible, and this makes a mass measurement possible.

As discussed by \citet{gould00} and in greater detail by \citet{lmc5},
$\pi_\e$ is actually the magnitude of a vector quantity, $\bpi_\e$,
whose direction is that of the lens-source relative motion.  We first
show that one component of $\bpi_\e$ is extremely well determined, so
that if its direction could also be constrained, $\pi_\e$ would also be
well determined.  Second, we show that the blended light for this event
is almost certainly the lens.  We outline how future space-based or
possibly ground-based observations could measure $\bmu_\rel$, the vector
lens-source relative proper motion \citep{han}.  When combined with the
very well determined $t_\e$ for this event, this would yield $\theta_\e$
through equation~(\ref{eqn:tedef}).  At the same time, such a
proper-motion measurement would give the direction of motion and so
tightly constrain $\pi_\e$.

\section{Data
\label{sec:data}}

The event [(RA,Dec) = (18:06:34.68, $-$26:01:16.2),
$(l,b)=(4.859,-2.550)$]
was initially discovered by the Optical Gravitational Lens
Experiment (OGLE, \citealt{udalski94}) and was alerted to the community as
OGLE-2003-BLG-175 through the OGLE-III Early Warning System
(EWS, \citealt{udalski03}) on 2003 May 28.  It was independently
rediscovered by Microlensing Observations for Astrophysics
(MOA, \citealt{bond01}) and designated MOA-2003-BLG-45 on 2003 July 6.
It achieved peak magnification on HJD$'\equiv $HJD$-2450000 = 2863.1$
(2003 August 11).

Observations were carried out by four groups from a total of eight
observatories: OGLE from Chile, MOA from New Zealand, the Microlensing
Followup Network \citep[$\mu$FUN;][]{yoo} from Chile and Israel, and the
Probing Lensing Anomalies Network (PLANET, \citealt{albrow98}) from
Chile, Perth, South Africa and Tasmania.  OGLE made a total 178 $I$ band
observations from 2001 August 6 to 2003 November 10, of which 119 were
during the 2003 season, using the 1.3m Warsaw telescope at the Las
Campanas Observatory, Chile, which is operated by the Carnegie Institute
of Washington.  The exposures were 120 seconds and photometry was
obtained using difference image analysis \citep{wozniak00}.  MOA made a
total of 522 $I$ band observations from 2000 April 12 to 2003 November
4, of which 303 were during the 2003 season, using the 0.6 m Boller \&
Chivens telescope at Mt.\ John University Observatory in New Zealand.

$\mu$FUN monitoring of the event began on July 7. Observations were made
at the 1.3m (ex-2MASS) telescope at Cerro Tololo InterAmerican
Observatory in Chile using ANDICAM (DePoy et al. 2003) and at the Wise
1m telescope at Mitzpe Ramon in Israel using the Wise TeK 1K CCD camera.
At CTIO, there were 210 observations in $I$ band, from July 7 to October
29, and 11 observations in $V$, covering a similar period (July 9 --
November 5). Exposures were 5 minutes each. Observations at Wise
consisted of 12 in $I$ band, covering the period July 8 to August 12,
and 56 observations using a clear filter. The latter sampled the
lightcurve densely just after peak, from August 12 to August
15. Photometry for all $\mu$FUN observations was done using DoPHOT
(Schechter, Mateo \& Saha 1993).

PLANET observations of this event included: 52 observations in $I$ band
using the 0.9m telescope at CTIO, from August 11 to August 18; 80
observations in $I$ band using the 0.6m telescope at Perth Observatory
in Australia, from August 6 to November 2; 165 observations in $R$ band
using the Danish 1.54m telescope at La Silla, Chile, from June 4 to
September 1; 6 observations in $I$ band using the South African
Astronomical Observatory 1m telescope at Sutherland, South Africa, on
August 5; and 59 observations in $I$ band using the Canopus Observatory
1m telescope in Tasmania, from August 5 to September 21. The data
reduction was done with the PLANET pipeline using PSF fitting photometry
with DoPHOT.

In fitting the lightcurve, we iteratively renormalized errors to obtain
a $\chi^2$ per degree of freedom of unity and eliminated points that
were farther than 3$\sigma$ from the best fit. For the data sets (OGLE,
MOA, $\mu$FUN[Chile $I$, Chile $V$, Israel clear, Israel $I$], PLANET
[Chile Danish $R$, Chile CTIO $I$, Perth, South Africa, Tasmania]),
there were initially (178, 522, 210, 11, 56, 12, 165, 52, 80, 6, 59)
data points, of which (175, 515, 203, 11, 56, 12, 161, 51, 76, 5, 51)
were incorporated into the final fit, with corresponding renormalization
factors (1.48, 1.179, 1.10, 1.00, 0.94, 0.81, 2.26, 1.06, 1.06, 2.69,
1.62).

\begin{figure}
\includegraphics[scale=0.65,angle=-90]{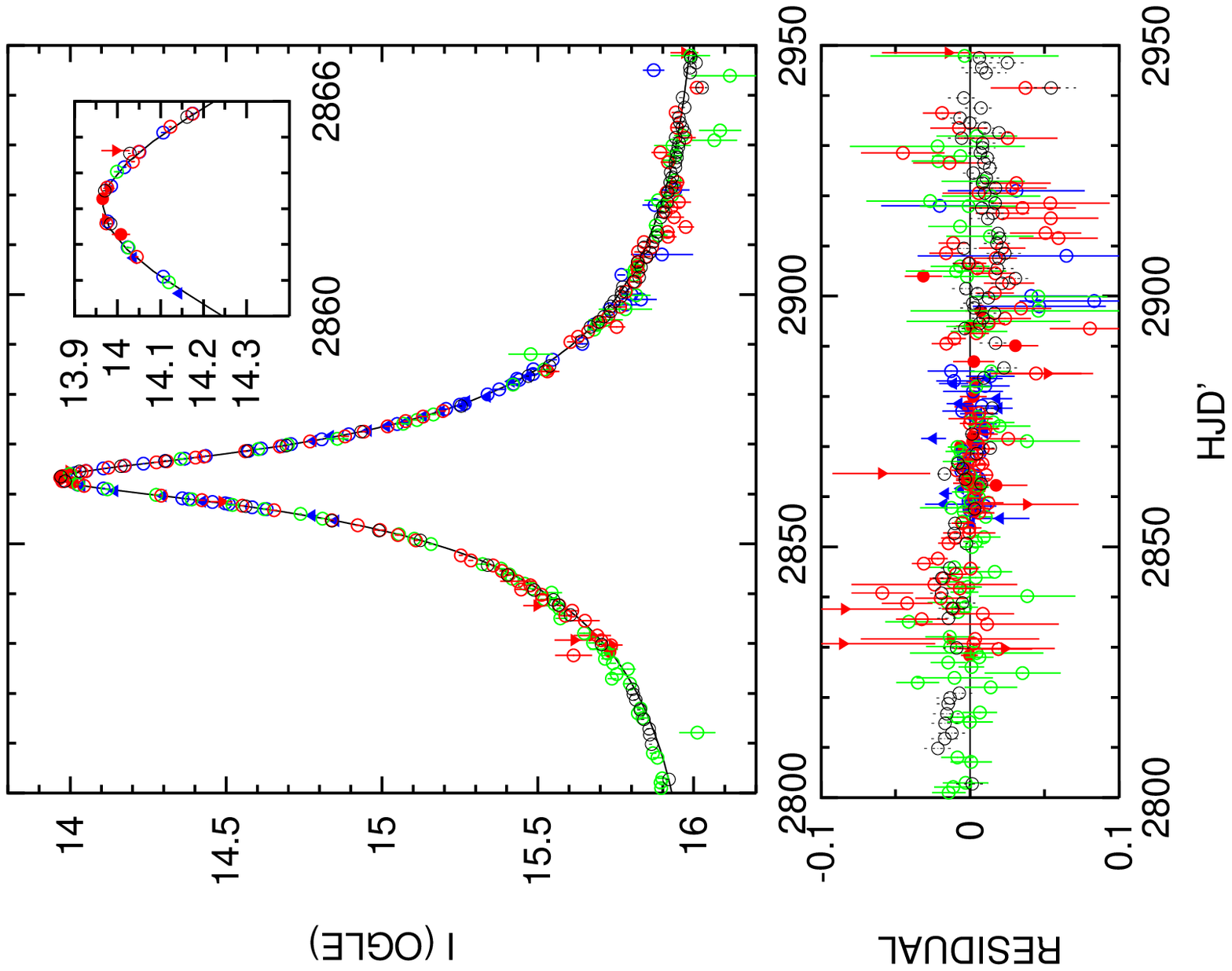}
\includegraphics[scale=0.65,angle=-90]{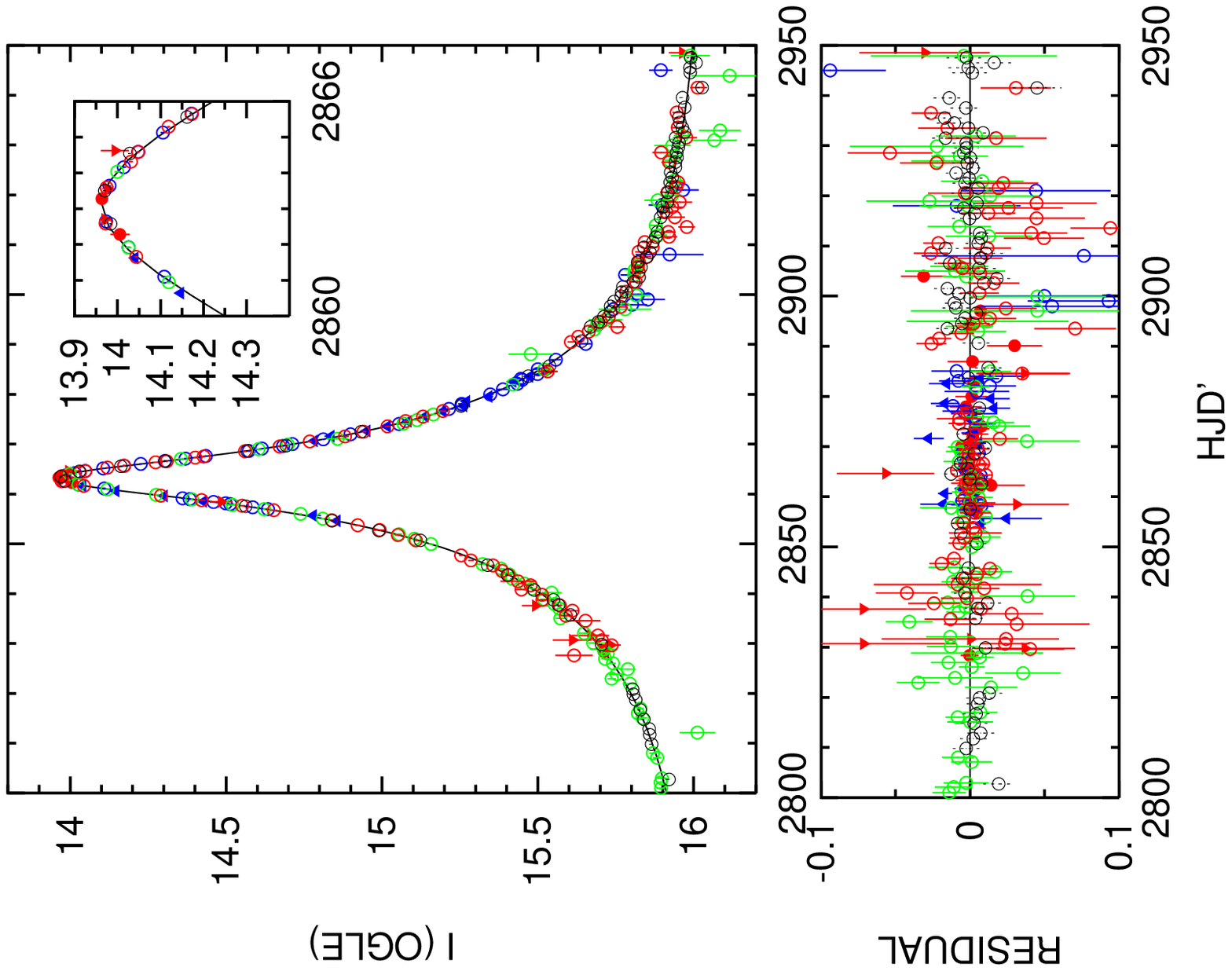}
\caption{Fit to the lightcurve in the nonparallax (left panel) and the
parallax (right panel) models. Data from different collaborations are
color-coded as follows: $\mu$FUN, red; MOA, green; OGLE, black; PLANET,
blue. Symbols indicate filter bands: $I$, open circles; $V$, inverted
triangles; {\it clear}, filled circles; $R$, upright triangles. Data
have been binned by night. The residuals from the nonparallax fit show
an asymmetry about the event peak; the asymmetry disappears when the
lightcurve is fit for parallax. The inset shows a zoomed-in view of the
peak.}
\label{fig:noplx}
\end{figure}

OGLE, MOA and $\mu$FUN photometric data for this event are publicly
available at \url{http://bulge.astro.princeton.edu/\string~ogle/},
\url{http://www.roe.ac.uk/\string~iab/alert/alert.html} and
\linebreak\url{http://www.astronomy.ohio-state.edu/\string~microfun/}.
The data from all four collaborations are shown in
Figure~\ref{fig:noplx} together with a standard fit to the lightcurve,
which shows strong residuals that are asymmetric about the peak.

\section{Lightcurve fitting
\label{sec:lc}}

All microlensing events are fit to the functional form
\begin{equation}
\label{eqn:foft}
F(t) = F_{\mathrm s} A(t) + F_{\mathrm b}
\end{equation}
where $F(t)$ is the observed flux, $F_{\mathrm s}$ is the source flux, which is
magnified by a factor $A(t)$, and $F_{\mathrm b}$ is the flux from any stars
blended with the source but not undergoing gravitational lensing.
For point-source point-lens events, $A(t) = A[u(t)]$, where $u$ is the
lens-source separation in units of $\theta_\e$  and \citep{pac86}
\begin{equation}
\label{eqn:aofu}
A(u) = \frac{u^2 + 2}{u{(u^2 + 4)}^{1/2}}\, .
\end{equation}

The event shows no significant signature of finite source effects,
implying that equations~(\ref{eqn:foft}) and (\ref{eqn:aofu}) are
appropriate.  However, it does show a highly significant asymmetry, of
the kind expected from parallax effects.  We therefore fit for five
geometric parameters (in addition to a pair of parameters, $F_{\mathrm
s}$ and $F_{\mathrm b}$, for each of the 11 observatory-filters
combinations).  Three of these five are the standard microlensing
parameters: the time of peak magnification, $t_0$, the Einstein crossing
time, $t_\mathrm{E}$, and the impact parameter $u_0 = u(t_0)$. The
remaining two are the microlens parallax $\boldsymbol\pi_\mathrm{E}$, a
vector whose magnitude gives the projected Einstein radius,
$\tilde{r}_\mathrm{E} \equiv \mathrm{AU}/\pi_\mathrm{E}$, and whose
direction gives the direction of lens-source relative motion.  We work
in the geocentric frame defined by \citet{lmc5}, so that the three
standard microlensing parameters ($t_0,t_{\rm E},u_0$) are nearly the
same as for the no-parallax fit (in which $\bpi_\e$ is fixed to be
zero). 

The parallactic distortion of the lightcurve has a component that is
asymmetric about the event peak, and one that is symmetric. The former
allows a determination of $\pi_{\e,\parallel}$, the component of the
parallax that is in the direction of the apparent acceleration of the
Sun projected onto the plane of the sky at event peak. The
symmetric distortion allows a determination of $\pi_{\e,\perp}$, the
component perpendicular to $\pi_{\e,\parallel}$. The direction of
$\pi_{\e,\perp}$ is chosen so that $(\pi_{\e,\parallel},\pi_{\e,\perp})$
form a right-handed coordinate system.
We fit for $\boldsymbol\pi_\mathrm{E}$, however, as
$\pi_{\mathrm{E},N}$ and $\pi_{\mathrm{E},E}$, the projections in the
North and East directions (in the equatorial coordinate system),
respectively. The error ellipse for these two $\bpi_\e$ parameters is
highly elongated. To quantify this effect, we also calculate
$(\pi_{\mathrm{E,1}}, \pi_{\mathrm{E,2}})$, the principal components of
$\bpi_\e$, as well as the position angle $\psi$ (north through east) of
the minor axis of the error ellipse.  The best-fit values thus obtained
are shown in Table~\ref{tab:fits}. However, microlensing lightcurve fits
can suffer from several degeneracies.

\subsection{Degeneracies in the models
\label{sec:degen}}

Degeneracies arise when the source-lens-observer relative trajectory
deviates from uniform rectilinear motion but there is not enough
information
in the lightcurve to distinguish among multiple possible trajectories.
We consider three types of degeneracies in our fits.

\subsubsection{Constant-acceleration degeneracy}
\label{sec:conacc}
Since this is a relatively short-duration ($t_\mathrm{E} \sim 63\,$days)
event, the change in acceleration over this timescale is relatively
small, and the fit is susceptible to the degeneracy derived by
\citet{smp03} in the limit of constant acceleration.  In the geocentric
frame adopted in this paper, the additional solution is expected to have
$u_0' = -u_0$, with the remaining parameters very similar to those of
the original solution \citep{smp03,lmc5}.  That is, the lens passes on
the opposite side of the source but otherwise the new trajectory is very
similar to the old one.  Table~\ref{tab:fits} shows that this is indeed
the case.  Moreover, the two solutions have almost identical $\chi^2$.

\subsubsection{Jerk-parallax degeneracy}

\citet{lmc5} generalized the analysis of \citet{smp03} to include jerk
and found an additional degeneracy whose parameters can be predicted
analytically from the parameters of the original solution together
with the known acceleration and jerk of the Earth at $t_0$.  This
prediction
has been verified for both MACHO-LMC-5 \citep{lmc5} and MOA 2003-BLG-37
\citep{moa37}. We search for this potential alternate solution
in two ways.  First, we adopt a seed solution at the location predicted
by \citet{lmc5} and search for a local minimum of the $\chi^2$ surface
in the neighborhood of this seed.  Second, we evaluate $\chi^2$ over a
grid of points in the $\bpi_\e$ plane and search for any local minima.
Neither search yields an additional solution.  We note that for
MACHO-LMC-5 (with timescale $t_\e\sim 30\,$days) the two solutions have
nearly identical $\chi^2$, while for MOA 2003-BLG-37 (with $t_\e\sim
42\,$days) the second minimum is disfavored at $\Delta\chi^2\sim 7$.  It
may well be that for events as long as OGLE-2003-BLG-175/MOA-2003-BLG-45
($t_\e\sim 63\,$days), the degeneracy is lifted altogether.

\subsubsection{Xallarap}

If the source is a component of a binary, its Keplerian motion will also
generate acceleration in the source-lens-observer trajectory.  Like the
Earth's motion, this is describable by the 7 parameters of a binary
orbit.  However, unlike the Earth's orbit, the binary-orbit parameters
are not known a priori.  Hence, while a parallax fit requires just two
parameters, $\bpi_\e$, (basically the size of the Einstein ring and the
direction of the lens-source relative motion relative to the Earth's
orbit), a full xallarap fit requires seven.  This proliferation of free
parameters may seem daunting but can actually be turned into an
advantage in understanding the event: if the full xallarap fit yields
parameters that are inconsistent with the Earth's orbit, then this is
proof that xallarap (rather than parallax alone) is at work.  On the
other hand, if the xallarap fit parameters are consistent with the
Earth's orbit, this is evidence that parallax is the predominant
acceleration effect.  Of course, the latter inference depends on the
size of the errors: if the xallarap parameters are tightly constrained
and agree with the Earth's orbit, this would be powerful evidence.  If
the errors are very large, mere consistency by itself does not provide a
strong argument.

We make two simplifications in our test for xallarap. First, instead of
adding five parameters (to make the full seven), we consider a more
restricted class of xallarap models with circular orbits.  This
eliminates two parameters, the eccentricity and the position angle of
the apse vector.  Hence only three additional parameters are required:
the inclination, phase, and period. Second, rather than introduce
additional free parameters into the fit, we conduct a grid search.

We find that the data do not discriminate among the models very
well. There is a large region of parameter space (including the Earth's
parameters) that is consistent with the data at the 2$\sigma$
level. Only very short orbital periods, $P < 0.2$ yr are excluded.

This exercise shows that, at least for this event, it is impossible
to discriminate between parallax and xallarap from the lightcurve
data alone.  Hence, some other argument is needed to decide between
these two possible interpretations of the acceleration that is
detected in the lightcurve.

\section{Characteristics of the Blended Light
\label{sec:blend}}

We now argue that the blended light is most likely due to the lens.
The key argument is astrometric: by measuring the centroid shift
during the event, we show that the source and the blend are aligned
to high precision and that the chance of such an alignment (if the
blend were not associated with the event) is extremely small. In
addition, the position of the blend on the color-magnitude diagram (CMD)
shows it to be foreground disk star.

\subsection{Astrometry
\label{sec:astrometry}}

If we ignore the displacement of the positions of the images relative to
that of the source (as justified below), then the position of the
source-blend centroid of light, $\btheta_c$, is given by the
flux-weighted average of the positions of the source and blend%
\footnote{\citet{amg95} calculate the shift in the centroid in the case
of zero proper motion. Note that their equation (2) is in error and
should read $\Delta\mathbf{r}_c = \mathbf{r} (1-f) \left[ 1 -
\frac{1}{Af + (1-f)} \right]$.}:
\begin{equation}
\btheta_c[A(t)] = \frac{A F_{\mathrm s}[\btheta_{\mathrm s}
+\bmu_{\mathrm s}(t-t_*)] + F_{\mathrm b}[\btheta_{\mathrm b} +
\bmu_{\mathrm b}(t-t_*)]}{A F_{\mathrm s} + F_{\mathrm b}} -
\frac{A_* F_{\mathrm s}\btheta_{\mathrm s} + F_{\mathrm
b}\btheta_{\mathrm b}}{A_* F_{\mathrm s} + F_{\mathrm b}} +\btheta_*,
\label{eqn:thetac}
\end{equation}
where $\btheta_{\mathrm s}$ and $\btheta_{\mathrm b}$ are the positions
of the source and blend at some fiducial time $t_*$. $\bmu_{\mathrm s}$
and $\bmu_{\mathrm b}$ are the proper motions of the source and blend,
and $\btheta_*$ is the centroid position at $t_*$.  For the time of
maximum magnification ($t_*=t_0$, $A_*=A_\max$), this equation can be
rewritten
\begin{equation}
\btheta_c[A(t)] = (\btheta_{\mathrm b} - \btheta_{\mathrm s})Z(A)
             + (A-1)\bmu_s W(A,t) + \bmu_c W(A,t) + \btheta_0
\label{eqn:thetac2}
\end{equation}
where
$\btheta_0$ is the centroid position at $t_0$, $\bmu_c \equiv (\bmu_s + r\bmu_b)$,
\begin{equation}
Z(A)\equiv \frac{(A_\max - A)r}{(A_\max + r)(A + r)},
\label{eqn:zdef}
\end{equation}
and
\begin{equation}
W(A,t) \equiv \frac{t-t_0}{A + r},\qquad r\equiv \frac{F_{\mathrm b}}{F_{\mathrm s}}.
\label{eqn:wdef}
\end{equation}

\begin{figure}
\includegraphics[scale=0.6]{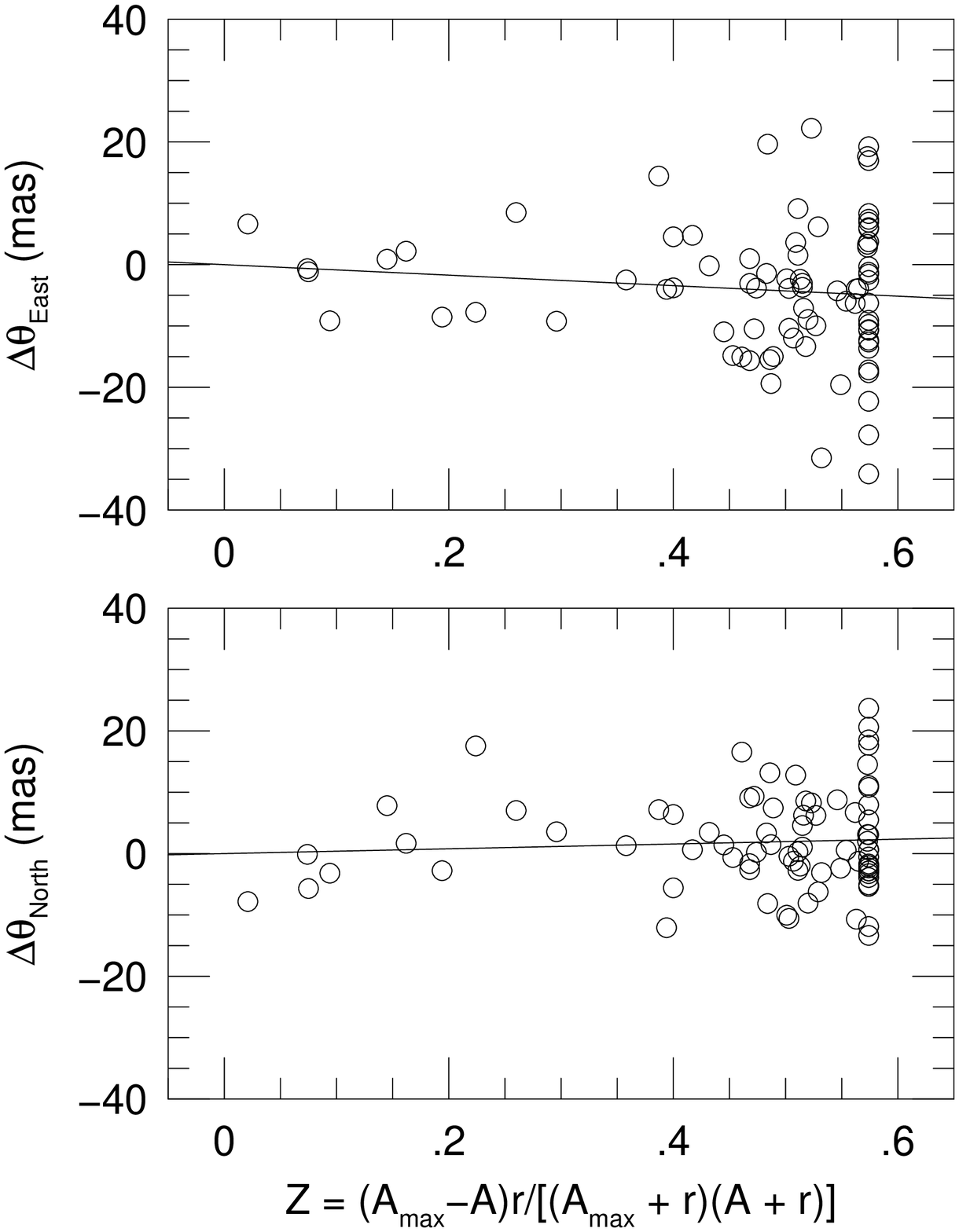}
\caption{Measurement of the separation of the microlensed source and the
unlensed blended light from OGLE astrometry.  As described in
\S~\ref{sec:astrometry}, the blend-source centroid should be at 
$\btheta_c[A(t)] = (\btheta_{\mathrm b}-\btheta_{\mathrm s})Z(A) +
(A\bmu_{\mathrm s} + r\bmu_{\mathrm b})W(A,t) + \btheta_0$
where $r=F_{\mathrm b}/F_{\mathrm s}$, $W(A,t) = (t-t_0)/(A+r)$, and
$Z(A)= (A_\max - A)r/[(A_\max + r)(A+r)]$.  The two panels
show the residuals, using the best-fit values of $\bmu_{\mathrm s}$,
$\bmu_{\mathrm b}$ and $\btheta_0$, in the East and North directions
but with the $(\btheta_{\mathrm b}-\btheta_{\mathrm s})Z(A)$ term removed. Hence
the slopes of the linear fits to these residuals, 
$(\theta_{\mathrm b}-\theta_{\mathrm s})_{\rm North}=3.9\pm 7.6\,$mas and
$(\theta_{\mathrm b}-\theta_{\mathrm s})_{\rm East}=-8.5\pm 10.5\,$mas 
give estimates for the source-blend separation.  This
separation is consistent with zero within small errors.}
\label{fig:astrometry}
\end{figure}

We have introduced the parameter $\bmu_c$ instead of using $\bmu_b$
since the latter is highly correlated with $\bmu_s$ and the linear
combination can be better constrained.
For the OGLE data, $r\approx2$ (see Table~\ref{tab:fits}). The quantities $Z$,
$W$ and $A(t)$ are determined from the fit to the lightcurve. We fit the
light centroid obtained from astrometry of 81 OGLE images taken both
before and during the event to equation (\ref{eqn:thetac2}) and find,
\begin{equation}
(\theta_{\mathrm b} - \theta_{\mathrm s})_{\rm North} = 3.9\pm 7.6\,
{\rm mas},\qquad (\theta_{\mathrm b} - \theta_{\mathrm s})_{\rm East} =
-8.5\pm 10.5\, {\rm mas}.
\label{eqn:offset}
\end{equation}
The astrometric measurement errors of the individual points are assumed
to be equal.  Their amplitude is determined by forcing $\chi^2$ per
degree of freedom to be unity.  They are found to be 8 and 11 mas in the
North and East directions, respectively.  In
Figure~\ref{fig:astrometry}, we show this fit together with the data
points plotted as $\Delta\btheta$ versus $Z$, where
\begin{equation}
\Delta\btheta\equiv \btheta_c - \bmu_{\mathrm s} (A-1)W - \bmu_c W - \btheta_0.
\label{eqn:deltathetadef}
\end{equation}
Here the $\btheta_c$ are the measured positions, while $\bmu_{\mathrm
s}$, $\bmu_c$, and $\btheta_0$ are the
best fit parameters.

\begin{figure}
\plotone{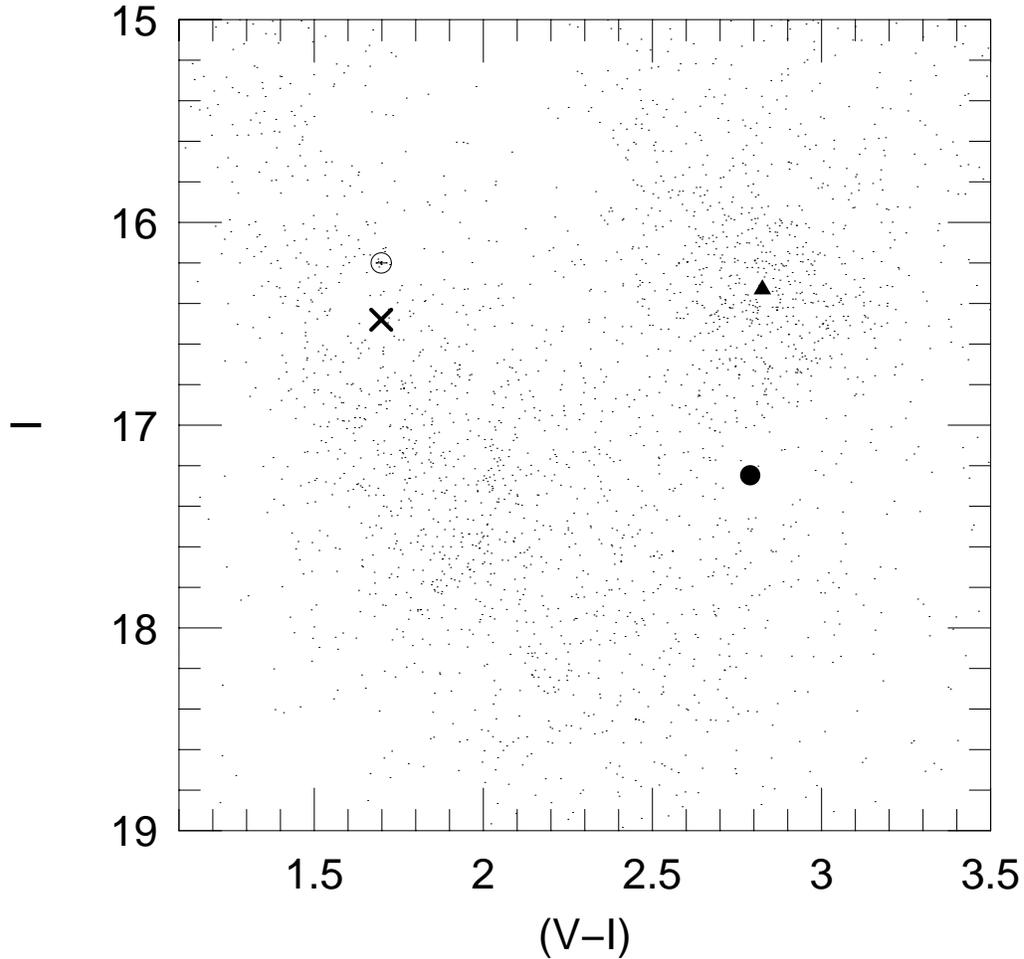}
\caption{Approximately calibrated color-magnitude diagram of stars within
a $6'$ square of OGLE-2003-BLG-175/MOA-2003-BLG-45.  See
\S~\ref{sec:mdblend} for details on the calibration procedure.  The
source ({\it filled circle}) lies a magnitude below the center of the
bulge clump ({\it triangle}) and therefore lies within or behind the
bulge.  The open circle shows the position of the blend based on
$\mu$FUN photometry, and the cross shows the position after the
correction based on OGLE photometry. The
blended light lies along the ``reddening sequence'' of foreground disk
main-sequence and turnoff stars, which suffer less extinction than the
bulge and therefore appear relatively bright and blue in the
diagram. Based on this diagram and arguments given in
\S~\ref{sec:mdblend}, we conclude that the blended light comes from a
star of mass $M_{\mathrm b}=0.75\,M_\odot$ and distance $D_{\mathrm b} =
1.1\,$kpc.}
\label{fig:cmd}
\end{figure}

Equation (\ref{eqn:offset}) shows that the source and blend have the
same position within about 15 mas.  There are only 42 stars/arcmin$^{2}$
in this field that are as bright or brighter than the blend.  The
probability that one of them would lie within 15 mas of the source is
therefore less than $10^{-5}$, unless the star were related to the
event.  If the blend is related to the event, there are only three
possibilities: (1) the blend is the lens, (2) the blend is a companion
to the lens, or (3) the blend is a companion to the source.  The last
possibility is ruled out by the color-magnitude diagram (CMD), which
shows that the blend lies in the foreground disk while the source lies
either in or behind the bulge (see Fig.~\ref{fig:cmd}).  While we cannot
immediately rule out that the blend is a companion to the lens rather
than the lens itself, we will show below that this hypothesis is
ultimately testable.  Moreover, even if the blend is a companion to the
lens, most of the arguments of this paper remain unaltered.  For the
moment, we ignore this possibility and tentatively assume that the blend
is the lens.

Before continuing, we note that neither of the proper-motion parameters
is determined with high precision.  We find
\begin{equation}
\label{eqn:mubs}
\bmu_{\mathrm s} = (-28,-46) \pm (64,89)\ \masyr
\quad \text{and} \quad (\bmu_{\mathrm
s} + r\bmu_{\mathrm b}) = (5,3)\pm (5,7)\ \masyr.
\end{equation}
  This means, in
particular, that the parameter of greatest physical interest
$(\bmu_{\mathrm b} -\bmu_{\mathrm s})$, which is a linear combination of
these two fit parameters, can only be determined with a precision of
about $100\,\masyr$, far larger than its plausible value.  Similarly,
the astrometric errors are at least an order of magnitude too large to
detect the motion of the image centroid relative to the source, which is
why we ignore it in this treatment.

\begin{figure}
\includegraphics[scale=0.44,angle=-90]{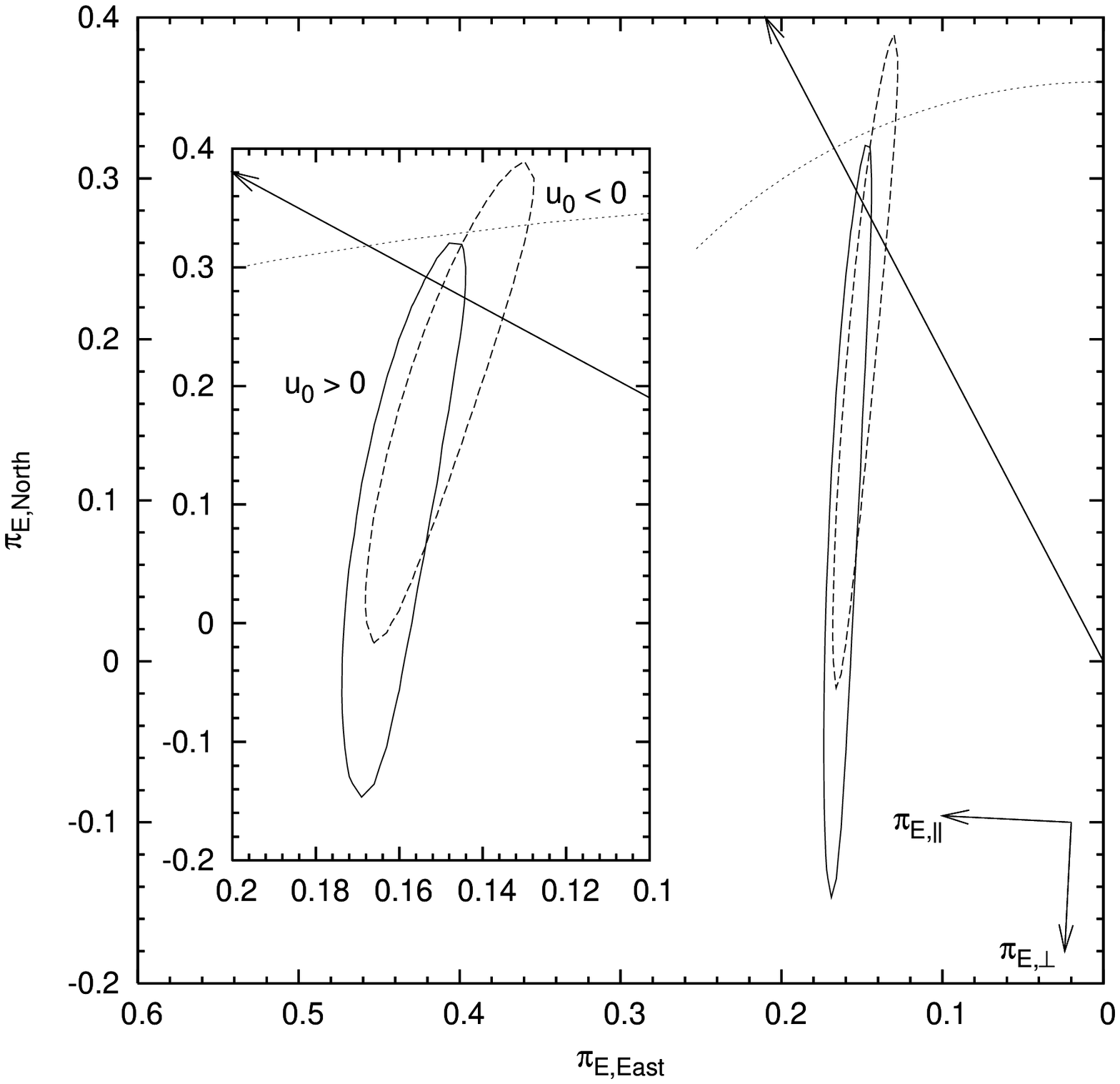}
\includegraphics[scale=0.44,angle=-90]{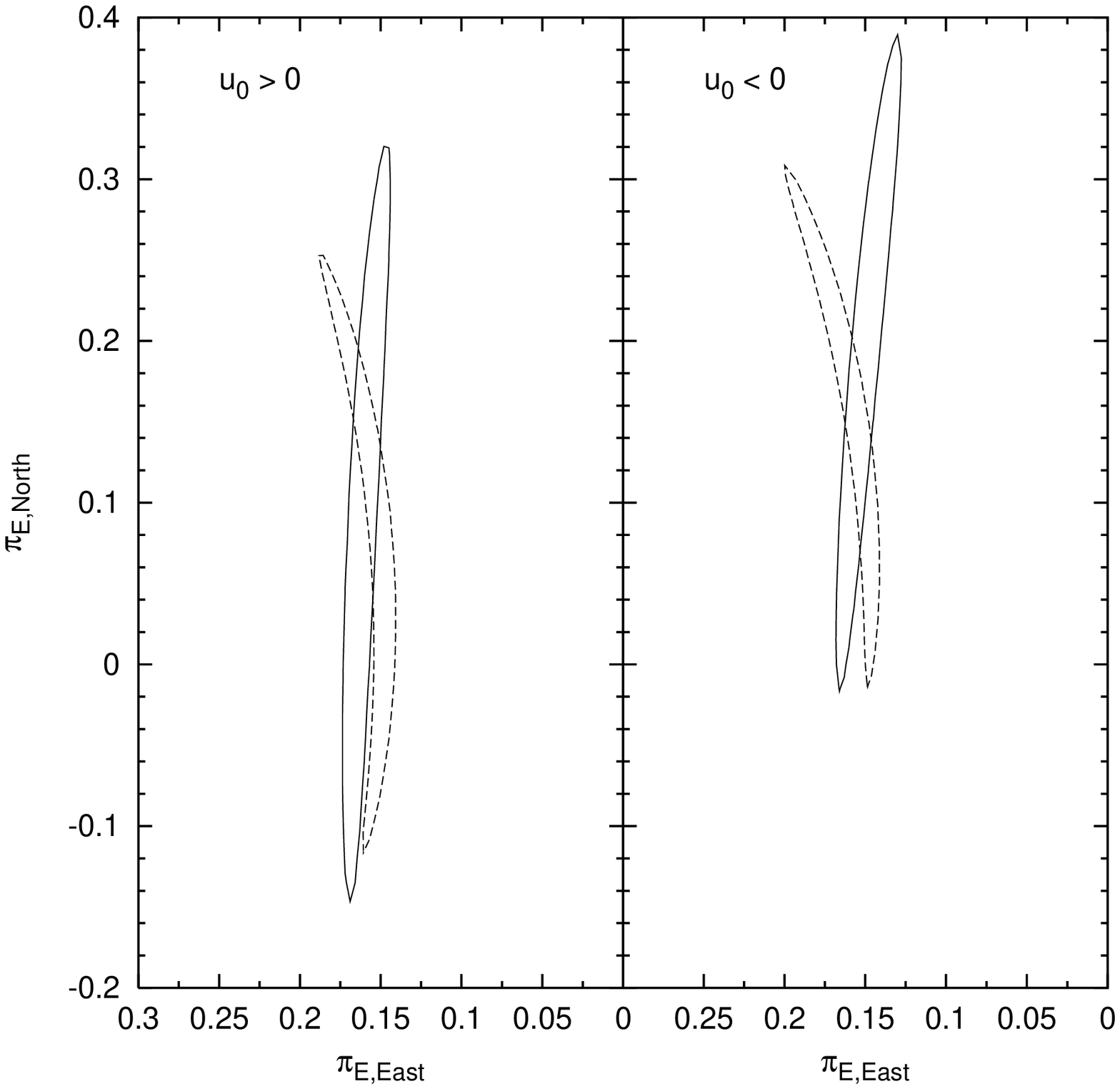}
\caption{\emph{Left panel:\/} $1\sigma$ contours of the $\chi^2$ surfaces, projected onto the
$(\pi_{\e,E},\pi_{\e,N})$ plane, for the $u_0 > 0$ and $u_0 < 0$
fits. The arrow shows the direction of Galactic rotation. The axis ratio
of the $u_0 > 0$ ``ellipse'' is $26:1$ and that of the $u_0 < 0$ one
$24:1$. %
Though only one component of $\bpi_\e$ is well-constrained, this figure
shows that the direction of lens-source relative proper motion is
consistent with the direction of Galactic rotation, which supports the
hypothesis that the lens is a foreground disk star. The dotted arcs are
parts of a circle of radius $\pi_\e=0.36$, the predicted microlens
parallax. The inset shows the contours with the horizontal scale
expanded. \emph{Right panel:\/} The solid contours are the same ones
shown in the left panel, and are in the geocentric frame. The contours
in dashed lines show them transformed to the heliocentric frame.}
\label{fig:piell}
\end{figure}

Finally, we note that if the blend is either the lens itself or a
companion to the lens, then one would expect the lens-source relative
motion to be in the direction of Galactic rotation (roughly North by
Northeast).  This is because the blend lies in the foreground disk while
the source lies in or behind the bulge. In fact, the parallax
measurement shows that the lens-source relative motion is consistent
with this direction (see Fig.~\ref{fig:piell}).

\subsection{Mass and Distance of the Blend
\label{sec:mdblend}}

Independent of whether the blend is indeed the lens, we can obtain a
rough estimate of the blend's mass and distance from its position on the
CMD by making use of disk color-magnitude relation of \citet{reid91},
$M_V = 2.89 + 3.37(V-I)$, together with the mass-luminosity relation of
\citet{allen}.  This estimate necessarily involves a number of
approximations. First, the two relations just mentioned have scatter in
them, which we ignore.  Second, while the reddening could in principle
be measured spectroscopically, no such measurement has been made.  We
therefore assume that the $I$-band extinction is related to the blend
distance by%
\footnote{Extinction is proportional to $\exp\left[-\frac{z}{z_h} +
\frac{R}{R_s}\right]$, where $z$ is the height above the Galactic plane,
$R$ is the distance, along the plane, from the Sun to the blend, and
$z_h$ and $R_s$ are the dust scale height and scale length,
respectively. $R$ has its origin at the Sun and \emph{increases} towards
the Galactic Center. Since $z=R\sin b$, for $z_h\sim 130$ pc, $R_s \sim
3$ kpc and $b = 2.5\degr$ the two terms inside the exponent almost
cancel.}$A_I=0.5\,(D_{\mathrm b}/\kpc)$.  Third, we must specify $R_{VI}
= A_V/E(V-I)$, the ratio of total to selective extinction.  This is
known to be anomalous toward the bulge, but while it varies somewhat
from one bulge line of sight to another, the measured values lie
consistently near $R_{VI}\sim 2.1$
\citep{pop00,ud03,sumi04}.  We therefore adopt this value.

Fourth, we must estimate the apparent color and magnitude of the blend.
While in principle the most straightforward step, under present
circumstances this is actually the most uncertain.  The flux of the
blend is a parameter of the fit to the lightcurve.  To determine a
color, we must have two such fluxes and so use the $\mu$FUN Chile
photometry since this is the only one of our observatories with data in
two photometric bands.  However, it is known that the $\mu$FUN Chile
photometry contains additional blended light relative to the OGLE
photometry.  First, the ratio of fit parameters, $r = F_{\mathrm
b}/F_{\mathrm s}$, is greater for $\mu$FUN Chile $I$ ($r\simeq 2.60$)
than for OGLE $I$ ($r \simeq 2.01$) despite the fact that the passbands
are very similar.  Second, OGLE photometry identifies additional sources
in the neighborhood of the source that $\mu$FUN photometry does not
identify and that therefore must be included in the $\mu$FUN blend.  If
the colors of these extra sources were known, they could just be removed
to find the color as well as the magnitude of the OGLE blend.
Unfortunately they are not known. For the purposes of this estimate we
assume the color of the blend is also the color of the lens. However,
the fact that the better-determined OGLE ratio $r\simeq2.01$ is less
than $r\simeq2.60$ for $\mu$FUN indicates that the lens may be fainter
by $0.28$ mag than its $\mu$FUN $I$ magnitude. We therefore use this
corrected value for our estimate.  Finally, the CMD has not been
directly calibrated to standard bands.  The OGLE $I$ fluxes are
calibrated to within a few tenths, and by identifying the OGLE and
$\mu$FUN $I$, we can therefore approximately calibrate the ordinate of
the $\mu$FUN CMD.  We then determine $A_I=2.0$ of the clump from the
calibrated $I$ of the clump and the known dereddened magnitude of the
clump, $I_0=14.32$ \citep{yoo}.  We then estimate $E(V-I)=1.82$ using
$R_{VI}=2.1$ and so from the known dereddened color of the clump,
$(V-I)_0=1.00$, calibrate the abscissa.  Clearly, the very complexity of
this approach as well as the sheer number of approximations leaves
something to be desired.  Nevertheless, as we are interested only in
rather crude mass and distance estimates for the blend, it will suffice.
We estimate,
\begin{equation}
I_{\mathrm b} = 16.48,\qquad (V-I)_{\mathrm b} = 1.67.
\label{eqn:blendcolmag}
\end{equation}

\begin{figure}
\includegraphics[scale=0.5,angle=-90]{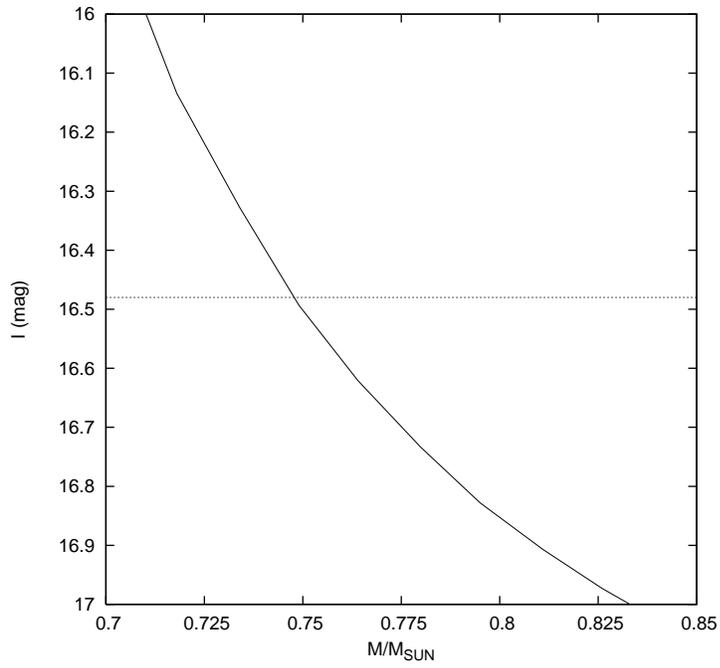}
\caption{Expected apparent $I$ magnitude vs.\ mass, as
explained in \S~\ref{sec:mdblend}. The observed $I =
16.48$ magnitude of the blend is marked with a dotted line. The
intersection of the two curves gives our estimate of the mass of the
blended light source, $M_{\mathrm b}=0.75\,M_\odot$.}
\label{fig:Im}
\end{figure}

To carry out our calculation, we consider trial stars as a function of
blend mass, $M_{\mathrm b}$.  For each mass, we obtain an absolute
magnitude $M_V$ using \citet{allen} and then a color
$(V-I)_{0,\mathrm{b}}$ using the \citet{reid91} color-magnitude
relation.  This gives a selective extinction $E(V-I) = (V-I)_{\mathrm b}
- (V-I)_{0,\mathrm{b}}$ and so an extinction $A_I=E(V-I)(R_{VI}-1)$, and
hence a distance $D_{\mathrm b} = A_I/(0.5\,{\rm mag/kpc})$. From this,
we obtain a predicted $I$ magnitude, $I=M_I+ 5\,\log D_{\mathrm
b}/10\,{\rm pc} + A_I$.  These predictions are shown in
Figure~\ref{fig:Im} where they are compared to the observed magnitude
$I_\mathrm{b}=16.48$.  From this comparison, we obtain
\begin{equation}
M_{\mathrm b}=0.75\,M_\odot, \qquad D_{\mathrm b}=1.1\,\kpc.
\label{eqn:massdist}
\end{equation}

\subsection{Microlens Parallax and Proper-Motion Predictions
\label{sec:parpmpred}}

If we identify the blend with the lens, and assume that the source lies
at $D_{\mathrm s}\sim 10\,\kpc$, we obtain $\pi_\rel = \pi_{\mathrm l} -
\pi_{\mathrm s} = 0.91\,\mas - 0.10\,\mas = 0.81\,\mas$. Assuming $M_{\mathrm l} =
M_{\mathrm b} = 0.75M_\sun$, and substituting these values into
equations~(\ref{eqn:thetaekappa}) and (\ref{eqn:pie}) yields,
\begin{equation}
\pi_\e = 0.36,\qquad \theta_\e = 2.2\,\mas.
\label{eqn:piethetaeval}
\end{equation}

We may now ask if these values, which are derived from the
photometrically-determined characteristics of the blend, are consistent
with what is known about the microlensing event. From Figure
\ref{fig:piell}, we see that the predicted parallax, $\pi_\e = 0.36$, is
consistent with the value observed at the $\sim1\,\sigma$ level. Combining
the Einstein radius $\theta_\e = 2.2\,\mas$ with the event's measured
timescale $t_\e = 63\,$days yields a proper motion
\begin{equation}
\mu_\rel = \frac{\theta_\e}{t_\e} = 13\,\masyr.
\label{eqn:murel}
\end{equation}

The only hard information we have on $\mu_\rel$ comes from the lack of
finite source effects, which puts a weak lower limit on the Einstein
radius, $\theta_\e > \theta_* / u_0$. Here $\theta_*$ is the source
radius and $u_0 = 0.05$ is the impact parameter. Using the standard
method to infer the angular source size from the instrumental CMD
\citep{yoo}, we find $\theta_* = 3.8 \mu$as. Hence $\theta_\e > 76 \mu$as,
and $\mu_\rel > \theta_\e/t_\e =
0.5\,\masyr$. Equation~(\ref{eqn:murel}) easily satisfies this
limit. However, the proper motion in equation~(\ref{eqn:murel}) is
somewhat higher than the typical ($\mu_\rel\sim 7\,\masyr$) proper motion
that would be expected for a disk lens moving with same rotation
velocity as the Sun and seen projected against a star with some random
motion in the bulge. But, given that the blend is so close ($D_{\mathrm
b}\sim 1.1\,\kpc$), the peculiar motions of the Sun and the blend
relative to the mean disk rotation may both contribute significantly to
$\mu_\rel$. Finally, the measurement of the blended motion $\bmu_c\equiv
\bmu_s + r\bmu_b = (5,3)\pm (5,7)\,\masyr$ (see eq.~[\ref{eqn:mubs}]),
also places indirect constraints on $\bmu_\rel= [\bmu_c-(r+1)\bmu_s]/r$.
That is, since $r\simeq 2$, equation~(\ref{eqn:murel}) implies $|3\bmu_s
- \bmu_c|=24\,\masyr$.  This constraint is not easily satisfied unless
either $\mu_s$ is anomalously fast or $\bmu_s$ is anti-aligned with
$\bmu_c$ (and so $\bmu_b$).  That is, $\bmu_s\cdot\bmu_c<0$.  However,
the latter option is quite plausible. The source could be retrogressing,
as would occur if it were in the far disk, and as would be consistent
with its position somewhat below the clump in the CMD. In that case, the
relative proper motion could be high without requiring rapid motion of the
centroid of light. Hence, the proper motion obtained in
equation~(\ref{eqn:murel}) is not unreasonable.

In brief, all the available evidence is consistent with the hypothesis
that the blend is the lens or a companion to the lens. In either case,
this opens the possibility that the lens mass can be precisely
determined by measuring the proper motion of the blend. There would
still remain the question of whether the mass that was measured was that
of the blend, in which case this measurement could be compared with more
accurate photometric and spectroscopic measurements than have been
obtained to date. We return to this question in \S~\ref{sec:lensblend}.

\subsection{Uncertainty in Lens Mass Estimates}
\label{sec:massunc}

The mass $M$ of the lens is given by
\begin{equation}
\label{eqn:mass}
M = \frac{\theta_\mathrm{E}}{\kappa \pi_\mathrm{E}} =
\frac{\mu_\mathrm{rel}t_\mathrm{E}}{\kappa \pi_\mathrm{E}}.
\end{equation}
The uncertainty in the mass estimate is therefore
\begin{equation}
\label{eq:uncmass}
\left(\frac{\sigma_M}{M}\right)^2 =
\left(\frac{\sigma_{\mu_\rel}}{\mu_\rel}\right)^2
+ \left(\frac{\sigma_{t_\e}}{t_\e}\right)^2 + \left(\frac{\sigma_{\pi_\e}}{\pi_\e}\right)^2.
\end{equation}

\begin{figure}
\includegraphics{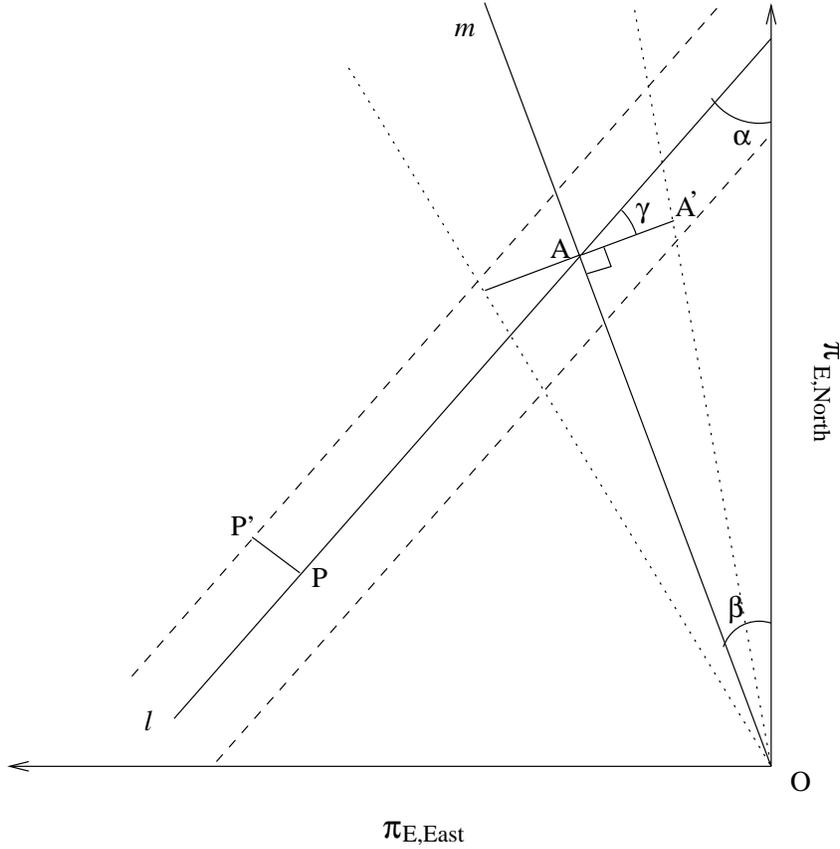}
\caption{
Even when only one component of the parallax $\bpi_\e$ is
well-determined, it may be possible to constrain the magnitude $\pi_\e$
if the source-lens relative proper motion $\bmu_\rel$ is measured. In
this figure it is assumed that the $(\pi_{\e,\parallel},\pi_{\e,\perp})$
axes are rotated from the
$(\pi_{\e,\mathrm{East}},\pi_{\e,\mathrm{North}})$ axes by an angle
$\alpha$, that $\pi_{\e,\parallel}$ is well-constrained and that
$\pi_{\e,\perp}$ is not constrained at all. The solid line $l$ is the
(very) long axis of the error ``ellipse'', which in this case is a strip
of width $2\overline{PP'} = 2\sigma_{\pi_{\e,\parallel}}$. The solid
line $m$ shows the direction of relative proper motion, and the dotted
lines on either side indicate the uncertainty in the direction. The
magnitude of $\bpi_\e$ corresponds to the length of $\overline{OA}$. The
uncertainty in the length of $\overline{OA}$ results from the
uncertainty $\overline{PP'}$ in $\pi_{\e,\parallel}$ and the uncertainty
$\overline{AA'}$ in the direction of $\bmu_\rel$.}
\label{fig:unc}
\end{figure}

In microlensing events in general, the fit to the lightcurve typically
produces a tight constraint on $t_\e$, so $\sigma_{t_\e} / t_\e$ is
usually small. Microlensing parallax is usually less well-determined. In
this event, for example, the fit constrains very well the component of
$\bpi_\e$ that is parallel to the direction of acceleration at the event
peak but the other component is only very poorly constrained. We refer
to these components%
\footnote{For short events, $t_\e \lesssim
\mathrm{yr}/2\pi$, the short axis of the error ellipse should line up
with the direction of acceleration \citep{gmb94} and this prediction has
been confirmed to high precision for two short events
\citep{moa37,jiang}. For OGLE-2003-BLG-175/MOA-2003-BLG-45 the
acceleration position angle is $87.1\degr$. So while the minor axis of
the error ellipse for the $u_0 > 0$ fit is aligned with the direction of
acceleration, that for the $u_0 < 0$ fit differs by $2.4\degr$. We ignore
this difference in this section, and refer to the principal components
as ``$\pi_{\e,\parallel}$'' and ``$\pi_{\e,\perp}$''.} %
as $\pi_{\e,\parallel}$ and $\pi_{\e,\perp}$. This situation is depicted
in Figure~\ref{fig:unc}, where the solid line $l$ shows the direction of
the long axis of the error ellipse for $\bpi_\e$ (see
Fig.~\ref{fig:piell} for comparison), and the dashed lines parallel to
$l$ and at a distance $\sigma_{\pi_{\e,\parallel}}$ indicate the
uncertainty. A hypothetical measurement of the direction of the relative
proper motion is shown in Figure~\ref{fig:unc} as the line $m$, with its
associated uncertainty indicated by the dotted lines on either side. The
magnitude of $\bpi_\e$ is then given by the distance from the origin to
the point of intersection of lines $l$ and $m$ ($\overline{OA}$ in
Fig.~\ref{fig:unc}).

The uncertainty in $\pi_\e$ thus has two contributions, one from the
width $\sigma_{\pi_{\e,\parallel}}$ and the other from
$\sigma_{\mu_\perp}$. The errors are not correlated and so may be added
in quadrature. The fractional uncertainty in $\pi_\e$ is therefore given
by
\begin{equation}
\label{eq:uncpi}
\left(\frac{\sigma_{\pi_\e}}{\pi_\e}\right)^2 =
\left(\frac{\sigma_{\pi_{\e,\parallel}}}{\pi_{\e,\parallel}}\sec\gamma\right)^2 + \left(\frac{\sigma_{\mu_\perp}}{\mu}\tan\gamma\right)^2
\end{equation}
where $\gamma = 90\degr - (\alpha + \beta)$ and the angles $\alpha$,
$\beta$ and $\gamma$ are as defined in Figure~\ref{fig:unc}.

For this event, the fit to the lightcurve gives
$(\sigma_{t_\e}/t_\e)=0.012$ and
$(\sigma_{\pi_{\e,\parallel}}/\pi_{\e,\parallel}) = 0.056$, both
small. The error in the mass will therefore be dominated by the
uncertainty in the proper motion measurement, unless it is very
accurate. If $(\sigma_{\mu_\rel}/\mu_\rel) =
(\sigma_{\mu_\perp}/\mu_\rel)$ and $(\alpha + \beta) \approx 45\degr$, a
10\% determination of $\mu_\rel$ would give a 16\% measurement of the
mass.

\section{Measuring the Proper Motion}

The actual determination of the lens mass depends on an accurate
measurement of the lens-source relative vector proper motion: the
magnitude is required to determine $\theta_\e = \mu_\rel t_\e$, while
the direction is required to determine $\pi_\e$ (see
\S~\ref{sec:massunc}).

There are two ways in which $\bmu_\rel$ may be observationally
determined. It will be possible, by waiting long enough, to resolve the
source and the blend into separate objects in ground-based
images. However, at the estimated rate of $13\,\masyr$ that might take
of the order of a decade.

The other method uses the higher resolution of space-based imaging and
the significantly different reddening of the source and the blend. The
blend is on the reddening sequence on the CMD, which indicates it lies
in front of most of the dust column to the source. The source, which
lies in or behind the bulge, is therefore much more reddened than the
blend. Hence, by imaging in a blue band such as $U$, in which the
source is expected to suffer high extinction, and in a red band such as
$I$, in which it suffers much less extinction, it would be possible to
measure the separation between the source and the blend from the offset
of their centroids in the two bands even before they are separately
resolved. The true separation $\Delta\btheta$ is given in terms of the
separation $\Delta\btheta_{UI}$ in the $U$ and $I$ band centroids by
\begin{equation}
\label{eq:centoff}
\Delta\btheta = 
\biggl[\frac{1}{1 + (F_{\mathrm s}/F_{\mathrm b})_U} - \frac{1}{1 + (F_{\mathrm s}/F_{\mathrm b})_I}\biggr]^{-1}
\Delta\btheta_{UI}
\end{equation}
where $(F_{\mathrm s}/F_{\mathrm b})_U$ and $(F_{\mathrm s}/F_{\mathrm b})_I$ are respectively the source/lens
flux ratios in $U$ and $I$. The proper motion is then simply $\bmu_\rel =
\Delta \btheta / \Delta t$, where $\Delta t$ is the time interval between
the event peak and the epoch of the observation. With the resolution of
the {\it Hubble Space Telescope}, such a measurement should be possible
3 years after the event peak. Note that while $(F_s/F_b)_I$ is known
from the microlensing event itself, the determination of $(F_s/F_b)_U$
requires a bit more work.  First, $(F_s+F_b)_U$ is measured directly in
the followup observations.  To find $F_{s,U}$, one should first note
that $(V-I)_s$ is well determined from the microlensing fit. Hence,
$(U-I)_s$ will be very similar to the $(U-I)$ of other clump stars with
the same $(V-I)$ as the source.  One can evaluate the error in this
determination from its scatter when applied to other clump stars.  This
procedure yields $F_{\mathrm s,U}$, and so (together with the total-flux
measurement) $(F_s/F_b)_U$.

Applying this method to blue and red plates from the 1950 Palomar Sky
Survey, we find a shift $\Delta\btheta_{BR} =
(0.30\pm0.17,-0.23\pm0.17)$ arcseconds, with the red centroid to the
north-east of the blue. While this direction agrees with the relative proper
motion expected for a disk lens, the magnitude is
consistent with zero and provides no useful constraint.

\subsection{Geocentric versus heliocentric frames}

\begin{figure}
\includegraphics[scale=0.7,angle=-90]{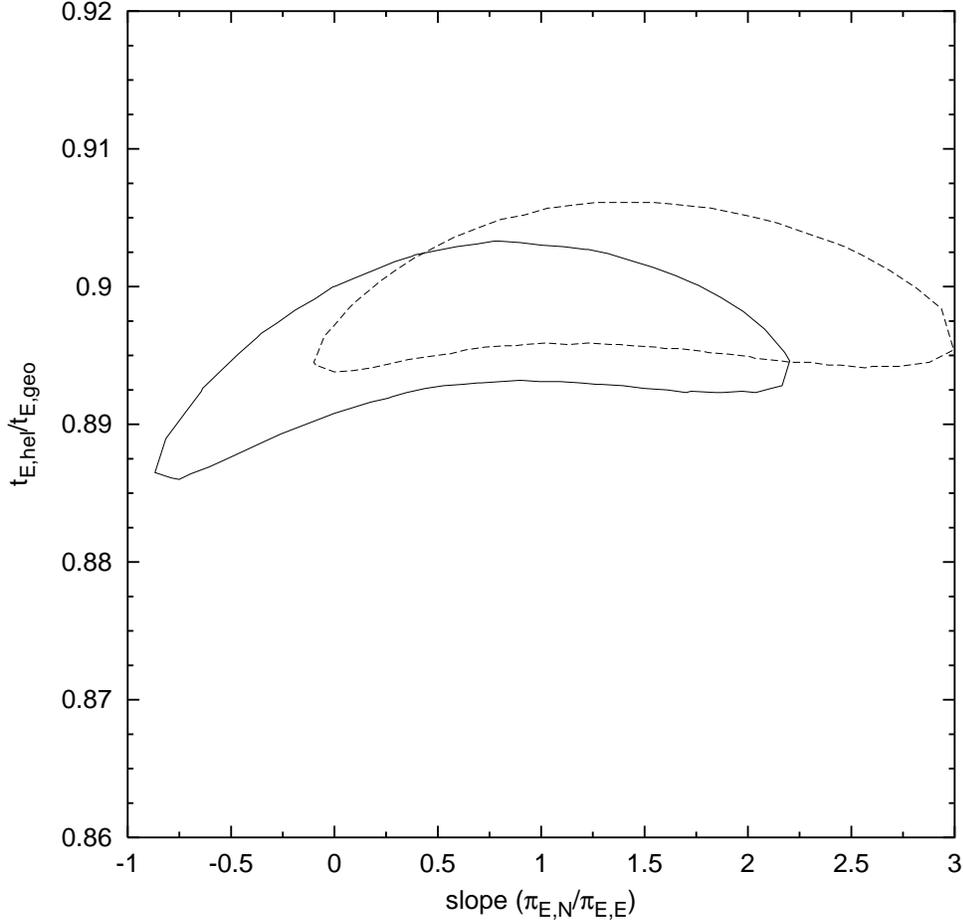}
\caption{Ratio of Einstein timescales $t_\e$ in the heliocentric and
 geocentric frames, evaluated along the contours shown in
Figure~\ref{fig:piell} for the $u_0 > 0$ (solid line) and the $u_0 < 0$
(dashed line) fits. For a given direction of lens-source relative proper motion
$\beta = \tan^{-1}(\pi_{\e,\mathrm{North}}/\pi_{\e,\mathrm{East}})$,
($\beta$ can be inferred from the direction of $\bmu_\rel$) the
width of the $1\sigma$ contours corresponds to a variation in
$t_{\e,\mathrm{hel}}/t_{\e,\mathrm{geo}}$ of about 3\%.}
\label{fig:tegeohel}
\end{figure}

Though we have been working in the geocentric frame in this paper, the
lens-source relative proper motion $\bmu_\rel$ is measured in
the heliocentric frame. Hence, to measure the mass $M = \mu_\rel
t_{\e,\mathrm{hel}}/\kappa \pi_\e$, the Einstein timescale $t_\e =
t_{\e,\mathrm{geo}}$ that we obtain from fitting the lightcurve must be
transformed to the heliocentric frame, in which it is not as well
determined. The geocentric and heliocentric timescales are related by
$\tilde{r}_\e = t_{\e,\mathrm{geo}} \tilde{v}_{\mathrm{geo}} =
t_{\e,\mathrm{hel}} \tilde{v}_\mathrm{hel}$, where $\tilde{r}_\e$ is the
projected Einstein radius and $\tilde{\mathbf{v}}$ is the projected
velocity in the appropriate frame. The transformation of the projected
velocity to the heliocentric frame is accomplished using the known
geocentric velocity of the Sun at event peak:
$\tilde{\mathbf{v}}_\mathrm{hel} = \tilde{\mathbf{v}}_\mathrm{geo} -
\mathbf{v}_\sun$.  Since $\tilde{v} \propto 1/(\pi_\e t_\e)$ and $t_\e$
is well-determined in the geocentric frame, the uncertainty in
$\tilde{v}_\mathrm{geo}$ is dominated by the uncertainty in $\pi_\e$
shown by the $1\sigma$ contours in Figure~\ref{fig:piell}.  As shown in
Figure~\ref{fig:tegeohel}, the corresponding uncertainty in the ratio
$t_{\e,\mathrm{hel}}/t_{\e,\mathrm{geo}} =
\tilde{v}_\mathrm{geo}/\tilde{v}_\mathrm{hel}$ is about 3\%, almost independent of the
actual direction of the proper motion.

\section{Distinguishing the Lens and Blend Hypotheses}
\label{sec:lensblend}
If the mass of the lens is measured from the proper motion of
the blend, it will still not automatically be known whether
the lens is the blend or is a companion to it.  Here we show
that the proper-motion measurement itself can help distinguish
these hypotheses.

There is no sign of binarity in the well-sampled lightcurve. This
provides a lower limit to the binary separation if the lens system is a
wide binary, or an upper limit if the system is a close binary, as
follows.

At large separations, the companion would induce a caustic in the
magnification profile of full width $\ell = 4 q^{-1}d^{-2}$, where $q$
is the lens/blend mass ratio, $d\theta_\e$ is their separation, and
$\ell\theta_\e$ is the angular width of the caustic.  Since the source
clearly did not traverse a caustic, $\ell<2\sqrt{2}u_0$, and indeed,
detailed fitting would provide somewhat tighter constraints \citep[see
Fig.\ 1 of][]{gg97}.  Hence, $d>(q u_0/\surd 2)^{-1/2} = 5.3\,q^{-1/2}$.  From
equation~(\ref{eqn:piethetaeval}), the Einstein radius associated with
the blend is expected to be $\theta_\e=2.2\,$mas.  The lens Einstein
radius would be smaller by $q^{1/2}$.  Hence, the separation of the lens
from the putatively distinct luminous blend must have been at least
\begin{equation}
|\btheta_{\mathrm l} - \btheta_{\mathrm b}| = d
q^{1/2}\theta_{\e,\mathrm{b}} > 12\,\rm mas
\label{eqn:thetamin}
\end{equation}
at the peak of the event.

This separation is already of order the measurement errors
from the OGLE astrometry (see eq.~[\ref{eqn:offset}]). 
If future proper-motion measurements are taken at multiple epochs, they
should be able to determine whether the blend-source relative motion
points back to a common position at the time of the event, or whether
the two were separated by at least the lower limit from
equation~(\ref{eqn:thetamin}).  Such a measurement would therefore
be able to determine whether the blend was the lens, or a companion
to the lens.

If the system is a close binary (either one of the components is
unseen, or both are visible but unresolved), then $d < (q^{1/2} +
q^{-1/2}) (u_0/\surd 2)^{1/2}$. For small $q$ this limit can be
approximated as $d < 1/(5.3\,q^{1/2})$. Since $v^2 = M/dr_\e$, where $v$ is
the orbital velocity of the secondary, that implies $v > 1.6 q^{1/4}
(M/M_\sun)^{1/2} v_\earth$, where we have used $\theta_\e = 2.2$ mas and
$\pi_\mathrm{l} \sim 1$ kpc. The radial velocity of the primary is then
$qv \gtrsim 42q^{5/4}$ km s$^{-1}$. This should be detectable unless the
companion is substellar ($q \ll 0.1$).

\acknowledgments

Work at OSU was supported by grants AST 02-01266 from the NSF and NAG
5-10678 from NASA. A.G. acknowledges support by NASA through Hubble
Fellowship grant \#HST-HF-01158.01-A awarded by STScI, which is operated
by AURA, Inc., for NASA, under contract NAS 5-26555. B.S.G. was
supported by a Menzel Fellowship from the Harvard College Observatory.
C.H. was supported by the Astrophysical Research Center for the
Structure and Evolution of the Cosmos (ARCSEC$''$) of Korea Science \&
Engineering Foundation (KOSEF) through the Science Research Program (SRC)
program. The MOA project is supported by the Marsden Fund of New
Zealand, the Ministry of Education, Culture, Sports, Science and
Technology (MEXT) of Japan, and the Japan Society for the Promotion of
Science (JSPS). Partial support to the OGLE project was provided with
the NSF grant AST-0204908 and NASA grant NAG5-12212 to B.~Paczy\'nski
and the Polish KBN grant 2P03D02124 to A.\ Udalski. A.U., I.S. and
K.\.Z. also acknowledge support from the grant ``Subsydium
Profesorskie'' of the Foundation for Polish Science.  M.D. acknowledges
postdoctoral support on the PPARC rolling grant PPA/G/O/2001/00475.
The allocation of observing time by IJAF at the Danish 1.54m telescope
at La Silla and support for the observations by the Danish Natural
Science Research Council (SNF) is acknowledged.

\begin{deluxetable}{lrrrrr}
\tabletypesize{\footnotesize}
\tablewidth{0pt}
\tablecaption{Best-Fit Parameters}
\tablehead{\colhead{} & \multicolumn{2}{c}{$u_0 > 0$ fit} & \colhead{} &
\multicolumn{2}{c}{$u_0 < 0$ fit} \\
\cline{2-3} \cline{5-6} \\
\colhead{Parameter} & \colhead{Value} &\colhead{Uncertainty} & \colhead{}
&\colhead{Value} & \colhead{Uncertainty}}

\startdata
$t_0$(days)               &  2863.1116 & 0.0065 & & 2863.1119  & 0.0059 \\
$u_0$                     &  0.0546    & 0.0008 & & $-0.0547$  & 0.0008 \\
$t_{\mathrm E}${(days)}   &  62.7894   & 0.8874 & & 63.2196    & 1.1297 \\
$\pi_{\rm {E},N}$         &  0.1108    & 0.7296 & & 0.2124     & 0.3463 \\
$\pi_{\rm {E},E}$         &  0.1603    & 0.0385 & & 0.1483     & 0.0335 \\
$\pi_{{\rm E},1}$         &  0.1658    & 0.0090 & & 0.1674     & 0.0090 \\
$\pi_{{\rm E},2}$         & $-0.1024$  & 0.7306 & & $-0.1977$  & 0.3478 \\
$\psi$                    & $87.1\degr$& ---    & & $84.7\degr$& --- \\
$(F_b/F_s)_1$             & 2.0057     & 0.0020 & & 2.0042     & 0.0018 \\
$(F_b/F_s)_2$             & 2.5965     & 0.0027 & & 2.5950     & 0.0024 \\
$(F_b/F_s)_3$             & 2.6459     & 0.0029 & & 2.6439     & 0.0026 \\
$(F_b/F_s)_4$             & 3.4187     & 0.0071 & & 3.4172     & 0.0067 \\
$(F_b/F_s)_5$             & 2.7371     & 0.0109 & & 2.7335     & 0.0106 \\
$(F_b/F_s)_6$             & 2.8999     & 0.0048 & & 2.8974     & 0.0045 \\
$(F_b/F_s)_7$             & 3.9967     & 2.7194 & & 3.9964     & 2.7173 \\
$(F_b/F_s)_8$             & 2.6091     & 0.0030 & & 2.6069     & 0.0028 \\
$(F_b/F_s)_9$             & 7.0954     & 0.0317 & & 7.0921     & 0.0301 \\
$(F_b/F_s)_{10}$             & 4.2482     & 0.1047 & & 4.2433  & 0.1039 \\
$(F_b/F_s)_{11}$             & 3.4974     & 0.0050 & & 3.4946  & 0.0045 \\

$\chi^2$                  &  1296.6528 & ---    & & 1295.9659  & ---    \\
\enddata
\tablecomments{Observatory/filter combinations for the ratios
$(F_b/F_s)$: 1=OGLE $I$, 2=$\mu$FUN Chile $I$, 3=MOA $I$, 4=$\mu$FUN
Wise $I$, 5=PLANET CTIO $I$, 6=PLANET Perth $I$, 7=PLANET SAAO $I$,
8=PLANET Tasmania $I$, 9=$\mu$FUN Chile $V$, 10=$\mu$FUN Wise clear,
11=PLANET Danish $R$.} 
\label{tab:fits}
\end{deluxetable}

\end{document}